\documentclass[a4paper,12pt]{article}
\usepackage{amsthm,amsmath,amssymb,latexsym,bm}

\newtheorem{thm}{Theorem}[section]

\newtheorem{lem}[thm]{Lemma}

\newtheorem{rem}[thm]{Remark}
\numberwithin{equation}{section}

\newcommand{\wt}{\widetilde}
\newcommand{\wh}{\widehat}

\newcommand{\DynD}{
\begin{figure}
\begin{center}
	\begin{picture}(170,70)
		\put(13,8){\circle*{6}}\put(0,5){\small$0$}
		\put(15,10){\line(1,1){20}}
		\put(13,56){\circle*{6}}\put(0,53){\small$1$}
		\put(15,54){\line(1,-1){20}}
		\put(37,32){\circle{6}}\put(34,17){\small$2$}
		\put(40,32){\line(1,0){20}}
		\put(63,32){\circle*{6}}\put(60,17){\small$3$}
		\put(66,32){\line(1,0){15}}
		\multiput(81,32)(4,0){5}{\line(1,0){2}}
		\put(101,32){\line(1,0){15}}
		\put(119,32){\circle{6}}\put(113,17){\small$2n$}
		\put(121,30){\line(1,-1){20}}
		\put(143,8){\circle*{6}}\put(149,5){\small$2n+2$}
		\put(121,34){\line(1,1){20}}
		\put(143,56){\circle*{6}}\put(149,53){\small$2n+1$}
	\end{picture}
\caption{Gradation of $\mathfrak{g}(D^{(1)}_{2n+2})$ of type
$(1,1,0,1,0,\ldots,1,0,1,1)$}
\end{center}
\end{figure}
}
\newcommand{\DynE}{
\begin{figure}
\begin{center}
	\begin{picture}(110,70)
		\put(55,41){\circle{6}}\put(42,38){\small$6$}
		\put(55,18){\line(0,1){20}}
		\put(55,67){\circle*{6}}\put(42,64){\small$0$}
		\put(55,44){\line(0,1){20}}
		\put(29,15){\circle{6}}\put(26,0){\small$2$}
		\put(52,15){\line(-1,0){20}}
		\put(3,15){\circle*{6}}\put(0,0){\small$1$}
		\put(26,15){\line(-1,0){20}}
		\put(81,15){\circle{6}}\put(78,0){\small$4$}
		\put(58,15){\line(1,0){20}}
		\put(107,15){\circle*{6}}\put(104,0){\small$5$}
		\put(84,15){\line(1,0){20}}
		\put(55,15){\circle*{6}}\put(52,0){\small$3$}
	\end{picture}
\caption{Gradation of $\mathfrak{g}(E^{(1)}_6)$ of type $(1,1,0,1,0,1,0)$}
\end{center}
\end{figure}
}

\title{Coupled Painlev\'{e} VI system with $E^{(1)}_6$-symmetry}
\author{Kenta Fuji and Takao Suzuki\\
{\small Department of Mathematics, Kobe University}\\
{\small Rokko, Kobe 657-8501, Japan}\\
{\small E-mail: suzukit@math.kobe-u.ac.jp}}
\date{}

\begin{document}

\maketitle

\begin{abstract}
We present an new system of ordinary differential equations with affine Weyl
group symmetry of type $E^{(1)}_6$.
This system is expressed as a Hamiltonian system of sixth order with a coupled
Painlev\'{e} VI Hamiltonian.
\end{abstract}

\section*{Introduction}

The Painlev\'{e} equations $P_{\rm{J}}$ $(\rm{J}=\rm{I},\ldots,\rm{VI})$ are
ordinary differential equations of second order.
It is known that these $P_{\rm{J}}$ admit the following affine Weyl group
symmetries \cite{O1}:
\[\begin{array}{|c|c|c|c|c|c|}\hline
	P_{\rm{I}}& P_{\rm{II}}& P_{\rm{III}}& P_{\rm{IV}}& P_{\rm{V}}&
	P_{\rm{VI}}\\[4pt]\hline\hline
	\text{--}& A_1^{(1)}& A_1^{(1)}\oplus A_1^{(1)}& A_2^{(1)}& A_3^{(1)}&
	D_4^{(1)}\\[4pt]\hline
\end{array}\]

Several extensions of the Painlev\'{e} equations have been studied from the
viewpoint of affine Weyl group symmetry.
The Noumi-Yamada system is a generalization of $P_{\rm{II}}$, $P_{\rm{IV}}$
and $P_{\rm{V}}$ for $A^{(1)}_n$-symmetry \cite{NY1}.
The coupled Painlev\'{e} VI system with $D^{(1)}_{2n+2}$-symmetry is also
studied \cite{S}.
In this paper, we present an new system of ordinary differential equations
with $E^{(1)}_6$-symmetry.
Our system can be expressed as a Hamiltonian system of sixth order with a
coupled Painlev\'{e} VI Hamiltonian.

In order to obtain this system, we consider a similarity reduction of a
Drinfeld-Sokolov hierarchy of type $E^{(1)}_6$.
The Drinfeld-Sokolov hierarchies are extensions of the KdV (or mKdV)
hierarchy \cite{DS}.
It is known that their similarity reductions imply several Painlev\'{e}
systems \cite{AS,FS1,FS2,KIK,KK1,KK2}.

The Drinfeld-Sokolov hierarchies are characterized by graded Heisenberg
subalgebras of affine Lie algebras.
In a recent work \cite{FS1,FS2}, we choosed the graded Heisenberg subalgebra
of $\mathfrak{g}(D^{(1)}_{2n+2})$ of type $(1,1,0,1,0,\ldots,1,0,1,1)$ for a
derivation of the coupled Painlev\'{e} VI system with
$D^{(1)}_{2n+2}$-symmetry.
In this paper, we apply a similar method to the case of $E^{(1)}_6$ by
choosing the graded Heisenberg subalgebra of type $(1,1,0,1,0,1,0)$.
The hierarchy defined thus implies our new system by similarity reduction.
\DynD\DynE

This paper is organaized as follows.
In Section 1, we present an explicit formula of a coupled Painlev\'{e} VI
system with $E^{(1)}_6$-symmetry.
In Section 2, we recall the affine Lie algebra $\mathfrak{g}(E^{(1)}_6)$ and
its graded Heisenberg subalgebra of type $(1,1,0,1,0,1,0)$.
In Section 3, we formulate a similarity reduction of a Drinfeld-Sokolov
hierarchy of type $E^{(1)}_6$.
In Section 4, we derive the coupled Painlev\'{e} VI system from the similarity
reduction.

\section{Main Result}\label{Sec:Main}

The Painlev\'{e} equation $P_{\rm{VI}}$ can be expressed as the following
Hamiltonian system \cite{IKSY,O2}:
\[
	s(s-1)\frac{dq}{ds}=\frac{\partial H_{\rm{VI}}}{\partial p},\quad
	s(s-1)\frac{dp}{ds}=-\frac{\partial H_{\rm{VI}}}{\partial q},
\]
with the Hamiltonian
$H_{\rm{VI}}=H_{\rm{VI}}(p,q,s;\beta_0,\beta_1,\beta_3,\beta_4)$ defined by
\[\begin{split}
	H_{\rm{VI}} &= q(q-1)(q-s)p^2 - \{(\beta_1-1)q(q-1)\\
	&\quad +\beta_3q(q-s)+\beta_4(q-1)(q-s)\}p + \beta_2(\beta_0+\beta_2)q,
\end{split}\]
where $\beta_i$ $(i=0,\ldots,4)$ are complex parameters satisfying
\[
	\beta_0 + \beta_1 + 2\beta_2 + \beta_3 + \beta_4 = 1.
\]

We define a {\it coupled} Hamiltonian $H$ by
\begin{equation}\begin{split}\label{Sys:E6_Ham}
	H &= H_{\rm{VI}}(p_1,q_1,s;\alpha_3,1-\alpha_1-2\alpha_2-2\alpha_3,
	\alpha_1,\alpha_3)\\
	&\quad + H_{\rm{VI}}(p_2,q_2,s;\alpha_3,1-2\alpha_3-2\alpha_4-\alpha_5,
	\alpha_5,\alpha_3)\\
	&\quad + H_{\rm{VI}}(p_3,q_3,s;\alpha_3,1-\alpha_0-2\alpha_3-2\alpha_6,
	\alpha_0,\alpha_3)\\
	&\quad + \sum_{1\leq i<j\leq3}\left\{(q_i-1)p_i+\alpha_{2i}\right\}
	\left\{(q_j-1)p_j+\alpha_{2j}\right\}(q_iq_j+s),
\end{split}\end{equation}
where $\alpha_i$ $(i=0,\ldots,6)$ are complex parameters satisfying
\[
	\alpha_0 + \alpha_1 + 2\alpha_2 + 3\alpha_3 + 2\alpha_4 + \alpha_5
	+ 2\alpha_6 = 1.
\]
Note that these parameters correspond to the simple roots of type $E^{(1)}_6$.
We consider a Hamiltonian system with Hamiltonian \eqref{Sys:E6_Ham}
\begin{equation}\label{Sys:E6}
	s(s-1)\frac{dq_i}{ds}=\{H,q_i\},\quad
	s(s-1)\frac{dp_i}{ds}=\{H,p_i\}\quad (i=1,2,3),
\end{equation}
where $\{\cdot,\cdot\}$ stands for the Poisson bracket defined by
\[
	\{p_i,q_j\} = \delta_{i,j},\quad \{p_i,p_j\} = \{q_i,q_j\} = 0\quad
	(i,j=1,2,3).
\]

The affine Weyl group $W(E^{(1)}_6)$ is generated by the transformations $r_i$
$(i=0,\ldots,6)$ acting on the simple roots as
\[
	r_i(\alpha_j) = \alpha_j - a_{ij}\alpha_i\quad (i,j=0,\ldots,6),
\]
where $A=(a_{ij})_{i,j=0}^6$ is the generalized Cartan matrix of type
$E^{(1)}_6$ defined by
\[
	A = \begin{bmatrix}
		2& 0& 0& 0& 0& 0& -1\\
		0& 2& -1& 0& 0& 0& 0\\
		0& -1& 2& -1& 0& 0& 0\\
		0& 0& -1& 2& -1& 0& -1\\
		0& 0& 0& -1& 2& -1& 0\\
		0& 0& 0& 0& -1& 2& 0\\
		-1& 0& 0& -1& 0& 0& 2\\
	\end{bmatrix}.
\]
Let $\pi_i$ $(i=1,2)$ be Dynkin diagram automorphisms acting on the simple
roots as
\[
	\pi_i(\alpha_j) = \alpha_{\sigma_i(j)}\quad (i=1,2; j=0,\ldots,6),
\]
where $\sigma_i$ $(i=1,2)$ are permutations defined by
\[
	\sigma_1=(01)(26),\quad \sigma_2=(05)(46).
\]
We consider an extension of $W(E^{(1)}_6)$
\[
	\wt{W} = \langle r_0,r_1,r_2,r_3,r_4,r_5,r_6,\pi_1,\pi_2\rangle,
\]
with the fundamental relations
\[\begin{array}{ll}
	r_i^2=1& (i=0,\ldots,6),\\[4pt]
	(r_ir_j)^{2-a_{ij}}=0& (i,j=0,\ldots,6; i\neq j),\\[4pt]
	\pi_i^2=1& (i=1,2),\\[4pt]
	(\pi_1\pi_2)^3=1,\\[4pt]
	\pi_ir_j=r_{\sigma_i(j)}\pi_i& (i=1,2; j=0,\ldots,6).
\end{array}\]

The action of the group $\wt{W}$ can be lifted to canonical transformations of
the Hamiltonian system \eqref{Sys:E6}.
Denoting by
\[\begin{array}{llll}
	\varphi_0=q_3-1,& \varphi_1=q_1-1,& \varphi_2=p_1,&
	\varphi_3=q_1q_2q_3-s,\\[4pt]
	\varphi_4=p_2,& \varphi_5=q_2-1,& \varphi_6=p_3.
\end{array}\]
we obtain
\begin{thm}\label{Bac_Trf}
The system \eqref{Sys:E6} with \eqref{Sys:E6_Ham} is invariant under the
action of birational canonical transformations $r_i$ $(i=0,\ldots,6)$ and
$\pi_i$ $(i=1,2)$ defined by
\[
	r_i(\alpha_j) = \alpha_j - a_{ij}\alpha_i,\quad
	r_i(\varphi_j) = \varphi_j + \frac{\alpha_i}{\varphi_i}
	\{\varphi_i,\varphi_j\}\quad (i,j=0,\ldots,6),
\]
and
\[
	\pi_i(\alpha_j) = \alpha_{\sigma_i(j)},\quad
	\pi_i(\varphi_j) = \varphi_{\sigma_i(j)}\quad (i=1,2; j=0,\ldots,6).
\]
\end{thm}

\section{Affine Lie algebra}\label{Sec:AffLie}

Following the notation of \cite{Kac}, we recall the affine Lie algebra
$\mathfrak{g}=\mathfrak{g}(E^{(1)}_6)$ and its graded Heisenberg subalgebra of
type $(1,1,0,1,0,1,0)$.

The affine Lie algebra $\mathfrak{g}$ is generated by the Chevalley generators
$e_i$, $f_i$, $\alpha_i^{\vee}$ $(i=0,\ldots,6)$ and the scaling element $d$
with the fundamental relations
\[\begin{split}
	&(\mathrm{ad}e_i)^{1-a_{ij}}(e_j)=0,\quad
	(\mathrm{ad}f_i)^{1-a_{ij}}(f_j)=0\quad (i\neq j),\\
	&[\alpha_i^{\vee},\alpha_j^{\vee}]=0,\quad
	[\alpha_i^{\vee},e_j]=a_{ij}e_j,\quad
	[\alpha_i^{\vee},f_j]=-a_{ij}f_j,\quad
	[e_i,f_j]=\delta_{i,j}\alpha_i^{\vee},\\
	&[d,\alpha_i^{\vee}]=0,\quad [d,e_i]=\delta_{i,0}e_0,\quad
	[d,f_i]=-\delta_{i,0}f_0,
\end{split}\]
for $i,j=0,\ldots,6$.
We denote the Cartan subalgebra of $\mathfrak{g}$ by
\[
	\mathfrak{h} = \bigoplus_{j=0}^{6}\mathbb{C}\alpha_j^{\vee}
	\oplus\mathbb{C}d.
\]
The canonical central element of $\mathfrak{g}$ is given by
\[
	K = \alpha_0^{\vee} + \alpha_1^{\vee} + 2\alpha_2^{\vee}
	+ 3\alpha_3^{\vee} + 2\alpha_4^{\vee} + \alpha_5^{\vee}
	+ 2\alpha_6^{\vee}.
\]
The normalized invariant form
$(\,|\,):\mathfrak{g}\times\mathfrak{g}\to\mathbb{C}$ is determined by the
conditions
\[\begin{array}{lll}
	(\alpha_i^{\vee}|\alpha_j^{\vee}) = a_{ij},& (e_i|f_j) = \delta_{i,j},&
	(\alpha_i^{\vee}|e_j) = (\alpha_i^{\vee}|f_j) = 0,\\[4pt]
	(d|d) = 0,& (d|\alpha_j^{\vee}) = \delta_{0,j},& (d|e_j) = (d|f_j) = 0,
\end{array}\]
for $i,j=0,\ldots,6$.

Consider the gradation
$\mathfrak{g} = \bigoplus_{k\in\mathbb{Z}}\mathfrak{g}_k$ of type
$(1,1,0,1,0,1,0)$ by setting
\[\begin{array}{ll}
	\deg\mathfrak{h}=\deg e_i=\deg f_i=0& (i=2,4,6),\\[4pt]
	\deg e_i=1,\quad \deg f_i=-1& (i=0,1,3,5).
\end{array}\]
With an element $\vartheta\in\mathfrak{h}$ such that
\[\begin{array}{ll}
	(\vartheta|\alpha_i^{\vee}) = 0& (i=2,4,6),\\[4pt]
	(\vartheta|\alpha_i^{\vee}) = 1& (i=0,1,3,5),
\end{array}\]
this gradation is defined by
\[
	\mathfrak{g}_k
	= \left\{x\in\mathfrak{g}\bigm|[\vartheta,x]=kx\right\}\quad
	(k\in\mathbb{Z}).
\]
We denote by
\[
	\mathfrak{g}_{<0} = \bigoplus_{k<0}\mathfrak{g}_{k},\quad
	\mathfrak{g}_{\geq0} = \bigoplus_{k\geq0}\mathfrak{g}_{k}.
\]

Such gradation implies the Heisenberg subalgebra of $\mathfrak{g}$
\[
	\mathfrak{s} = \{x\in\mathfrak{g}\bigm|[x,\Lambda_1]=\mathbb{C}K\},
\]
with an element of $\mathfrak{g}_1$
\[\begin{split}
	\Lambda_1 &= e_1 + 2e_3 + e_5 + e_{21} + e_{60} + e_{23}\\
	&\quad + e_{43} + e_{63} + e_{234} + e_{236} + e_{436} + 2e_{6234},
\end{split}\]
where
\[
	e_{i_1i_2\ldots i_nj}
	= \mathrm{ad}e_{i_1}\mathrm{ad}e_{i_2}\ldots\mathrm{ad}e_{i_n}(e_j).
\]
Note that $\mathfrak{s}$ admits the gradation of type $(1,1,0,1,0,1,0)$,
namely
\[
	\mathfrak{s} = \bigoplus_{k\in\mathbb{Z}}\mathfrak{s}_k,\quad
	\mathfrak{s}_k\subset\mathfrak{g}_k.
\]
We also remark that the positive part of $\mathfrak{s}$ has a graded bases
$\left\{\Lambda_k\right\}_{k=1}^{\infty}$ satisfying
\[
	[\Lambda_k,\Lambda_l] = 0,\quad
	[\vartheta,\Lambda_k] = n_k\Lambda_k\quad (k,l=1,2,\ldots),
\]
where $n_k$ stands for the degree of element $\Lambda_k$ defined by
\[\begin{array}{lll}
	n_{6l+1} = 6l+1,& n_{6l+2} = 6l+1,& n_{6l+3} = 6l+2,\\[4pt]
	n_{6l+4} = 6l+4,& n_{6l+5} = 6l+5,& n_{6l+6} = 6l+5.
\end{array}\]

We formulate the Drinfeld-Sokolov hierarchy of type $E^{(1)}_6$ associated
with the Heisenberg subalgebra $\mathfrak{s}$ by using these $\Lambda_k$ in
the next section.

\begin{rem}
The isomorphism classes of the Heisenberg subalgebras are in one-to-one
correspondence with the conjugacy classes of the finite Weyl group
{\rm\cite{KP}}.
In the notation of {\rm\cite{C}}, the Heisenberg subalgebra $\mathfrak{s}$
introduced above corresponds to the regular primitive conjugacy class
$E_6(a_2)$ of the Weyl group $W(E_6)${\rm;} see {\rm\cite{DF}}.
\end{rem}

\section{Drinfeld-Sokolov hierarchy}\label{Sec:DS}

In this section, we formulate a similarity reduction of a Drinfeld-Sokolov
hierarchy of type $E^{(1)}_6$ associated with the Heisenberg subalgebra
$\mathfrak{s}$.

In the following, we use the notation of infinite dimensional groups
\[
	G_{<0} = \exp(\wh{\mathfrak{g}}_{<0}),\quad
	G_{\geq0} = \exp(\wh{\mathfrak{g}}_{\geq0}),
\]
where $\wh{\mathfrak{g}}_{<0}$ and $\wh{\mathfrak{g}}_{\geq0}$ are completions
of $\mathfrak{g}_{<0}$ and $\mathfrak{g}_{\geq0}$ respectively.

Let $X(0)\in G_{<0}G_{\geq0}$.
Introducing the time variables $t_k$ $(k=1,2,\ldots)$, we condider a
$G_{<0}G_{\geq0}$-valued function
\[
	X = X(t_1,t_2,\ldots) 
	= \exp\left(\sum_{k=1,2,\ldots}t_k\Lambda_k\right)X(0).
\]
Then we have a system of partial differential equations
\begin{equation}\label{Eq:Gauss}
	X\partial_kX^{-1} = \partial_k- \Lambda_k\quad (k=1,2,\ldots),
\end{equation}
where $\partial_k=\partial/\partial t_k$, defined through the adjoint action
of $G_{<0}G_{\geq0}$ on $\wh{\mathfrak{g}}_{<0}\oplus\mathfrak{g}_{\geq0}$.
Via the decomposition
\[
	X = W^{-1}Z,\quad W\in G_{<0},\quad Z\in G_{\geq0},
\]
the system \eqref{Eq:Gauss} implies a system of partial differential equations
\begin{equation}\label{Eq:Sato}
	\partial_k - B_k = W(\partial_k-\Lambda_k)W^{-1}\quad (k=1,2,\ldots),
\end{equation}
where $B_k$ stands for the $\mathfrak{g}_{\geq0}$-component of
$W\Lambda_kW^{-1}\in\wh{\mathfrak{g}}_{<0}\oplus\mathfrak{g}_{\geq0}$.
The Zakharov-Shabat equations
\begin{equation}\label{ZS_DS}
	[\partial_k-B_k,\partial_l-B_l] = 0\quad (k,l=1,2,\ldots),
\end{equation}
follows from the system \eqref{Eq:Sato}.

Under the system \eqref{Eq:Sato}, we consider the operator
\[
	\mathcal{M} = W\exp\left(\sum_{k=1,2,\ldots}t_k\Lambda_k\right)\vartheta
	\exp\left(-\sum_{k=1,2,\ldots}t_k\Lambda_k\right)W^{-1}.
\]
Then the operator $\mathcal{M}$ satisfies
\begin{equation}\label{Sim_Red}
	[\partial_k-B_k,\mathcal{M}] = 0\quad (k=1,2,\ldots).
\end{equation}
Note that
\[
	\mathcal{M} = W\vartheta W^{-1}
	- \sum_{k=1,2,\ldots}n_kt_kW\Lambda_kW^{-1}.
\]

Now we require that the similarity condition
$\mathcal{M}\in\mathfrak{g}_{\geq0}$ is satisfied.
Then we have
\[
	\mathcal{M} = \vartheta - \sum_{k=1,2,\ldots}n_kt_kB_k.
\]
We also assume that $t_k=0$ for $k\geq3$.
Then the systems \eqref{ZS_DS} and \eqref{Sim_Red} are equivalent to
\begin{equation}\begin{split}\label{Sim_Red_2}
	&[\partial_1-B_1,\partial_2-B_2] = 0,\\
	&[\partial_k-B_k,\vartheta-t_1B_1-t_2B_2] = 0\quad (k=1,2).
\end{split}\end{equation}
We regard the system \eqref{Sim_Red_2} as a similarity reduction of
Drinfeld-Sokolov hierarchy of type $E^{(1)}_6$.

The $\mathfrak{g}_{\geq0}$-valued functions $B_k$ $(k=1,2)$ are expressed in
the form
\[
	B_k = U_k + \Lambda_k,\quad
	U_k = \sum_{i=0}^{6}u_{k,i}\alpha_i^{\vee} + \sum_{i=2,4,6}x_{k,i}e_i
	+ \sum_{i=2,4,6}y_{k,i}f_i.
\]
In terms of the operators $U_k\in\mathfrak{g}_0$, this similarity reduction
can be expressed as
\begin{equation}\begin{split}\label{Sim_Red_U}
	&\partial_1(U_2) - \partial_2(U_1) + [U_2,U_1] = 0,\\
	&[\Lambda_1,U_2] - [\Lambda_2,U_1] = 0,\\
	&t_1\partial_1(U_k) + t_2\partial_2(U_k) + U_k = 0\quad (k=1,2).
\end{split}\end{equation}
Note that the operators $\Lambda_k\in\mathfrak{g}_1$ are given by
\[\begin{split}
	\Lambda_1 &= e_1 + 2e_3 + e_5 + e_{21} + e_{60} + e_{23}\\
	&\quad + e_{43} + e_{63} + e_{234} + e_{236} + e_{436} + 2e_{6234},\\
	\Lambda_2 &= 2e_0 - 2e_3 - 2e_5 - 2e_{21} - 2e_{45} + 2e_{23}\\
	&\quad + 2e_{43} - 7e_{63} - 4e_{234} + 5e_{236}
	- 4e_{436} - 2e_{6234}.
\end{split}\]

In the following, we use the notation of a $\mathfrak{g}_{\geq0}$-valued
1-form $\mathcal{B}=B_1dt_1+B_2dt_2$ with respect to the coordinates
$\bm{t}=(t_1,t_2)$.
Then the similarity reduction \eqref{Sim_Red_2} is expressed as
\begin{equation}\label{Sim_Red_ED}
	d_{\bm{t}}\mathcal{M} = [\mathcal{B},\mathcal{M}],\quad
	d_{\bm{t}}\mathcal{B} = \mathcal{B}\wedge\mathcal{B},
\end{equation}
where $d_{\bm{t}}$ stands for an exterior differentation with respect to
$\bm{t}$.
Denoting by
\[
	\mathcal{M}_1 = -t_1\Lambda_1 - t_2\Lambda_2,\quad
	\mathcal{B}_1 = \Lambda_1dt_1 + \Lambda_2dt_2,
\]
we can express the operators $\mathcal{M}$ and $\mathcal{B}$ in the form
\[\begin{split}
	\mathcal{M} &= \theta + \sum_{i=2,4,6}\xi_ie_i
	+ \sum_{i=2,4,6}\psi_if_i + \mathcal{M}_1,\\
	\mathcal{B} &= \bm{u} + \sum_{i=2,4,6}\bm{x}_ie_i
	+ \sum_{i=2,4,6}\bm{y}_if_i + \mathcal{B}_1,
\end{split}\]
where
\[
	\theta = \vartheta + \sum_{i=0}^6\theta_i\alpha^{\vee}_i,\quad
	\bm{u} = \sum_{i=0}^6\bm{u}_i\alpha^{\vee}_i.
\]
The system \eqref{Sim_Red_ED} is expressed in terms of these variables as
follows:
\[\begin{split}
	&d_{\bm{t}}\theta_i = \bm{x}_i\psi_i - \bm{y}_i\xi_i,\quad
	d_{\bm{t}}\theta_j = 0,\\
	&d_{\bm{t}}\xi_i = (\bm{u}|\alpha^{\vee}_i)\xi_i
	- \bm{x}_i(\theta|\alpha^{\vee}_i),\\
	&d_{\bm{t}}\psi_i = -(\bm{u}|\alpha^{\vee}_i)\psi_i
	+ \bm{y}_i(\theta|\alpha^{\vee}_i),
\end{split}\]
and
\[\begin{split}
	&d_{\bm{t}}\bm{u}_i = \bm{x}_i\wedge\bm{y}_i
	+ \bm{y}_i\wedge\bm{x}_i,\quad d_{\bm{t}}\bm{u}_j = 0,\\
	&d_{\bm{t}}\bm{x}_i = (\bm{u}|\alpha^{\vee}_i)\wedge\bm{x}_i,\quad
	d_{\bm{t}}\bm{y}_i = -(\bm{u}|\alpha^{\vee}_i)\wedge\bm{y}_i,
\end{split}\]
for $i=2,4,6$ and $j=0,1,3,5$.

In this section, we proposed three representations \eqref{Sim_Red_2},
\eqref{Sim_Red_U} and \eqref{Sim_Red_ED} of the similarity reduction.
In the following, we use the system \eqref{Sim_Red_ED} in order to derive the
system \eqref{Sys:E6}.

\section{Derivation of Coupled $P_{\rm{VI}}$}\label{Sec:CP6}

In this section, we derive the Hamiltonian system \eqref{Sys:E6} from the
similarity reduction \eqref{Sim_Red_ED}.
Let $\mathfrak{n}_{+}$ be the subalgebra of $\mathfrak{g}$ generated by $e_i$
$(i=0,\ldots,6)$ and $\mathfrak{b}_{+}=\mathfrak{h}\oplus\mathfrak{n}_{+}$
the Borel subalgebra of $\mathfrak{g}$.
We introduce below a gauge transformation for the system
\eqref{Sim_Red_ED}
\[
	\mathcal{M}^{+} = \exp(\mathrm{ad}(\Gamma))\mathcal{M},\quad
	d_{\bm{t}}-\mathcal{B}^{+}
	= \exp(\mathrm{ad}(\Gamma))(d_{\bm{t}}-\mathcal{B}),
\]
with $\Gamma\in\mathfrak{g}_0$ such that $\mathcal{M}^{+}$ and
$\mathcal{B}^{+}$ should take values in $\mathfrak{b}_{+}$.

We first consider a gauge transformation
\[
	\mathcal{M}^* = \exp(\mathrm{ad}(\Gamma_1))\mathcal{M},\quad
	d_{\bm{t}}-\mathcal{B}^*
	= \exp(\mathrm{ad}(\Gamma_1))(d_{\bm{t}}-\mathcal{B}),
\]
with $\Gamma_1\in\mathfrak{g}_0\cap\mathfrak{b}_{+}$ such that
\[
	\exp(\mathrm{ad}(\Gamma_1))(\mathcal{M}_1)
	= \sum_{i=0,1,3,5}c_ie_i + e_{21} + e_{45} + e_{60} + e_{23} + e_{43}
	+ c_{63}e_{63} + e_{234},
\]
Note that $c_0$, $c_1$, $c_3$, $c_5$ and $c_{63}$ are algebraic functions in
$t_1$ and $t_2$.
Then We have
\begin{equation}\label{Sim_Red_ED_2}
	d_{\bm{t}}\mathcal{M}^* = [\mathcal{B}^*,\mathcal{M}^*],\quad
	d_{\bm{t}}\mathcal{B}^* = \mathcal{B}^*\wedge\mathcal{B}^*.
\end{equation}
With the notation
\[
	\mathcal{M}_1^*=\exp(\mathrm{ad}(\Gamma_1))(\mathcal{M}_1),\quad
	\mathcal{B}_1^*=\exp(\mathrm{ad}(\Gamma_1))(\mathcal{B}_1),
\]
the operators $\mathcal{M}^*$ and $\mathcal{B}^*$ are expressed in the form
\[\begin{split}
	\mathcal{M}^* &= \theta^* + \sum_{i=2,4,6}\xi^*_ie_i
	+ \sum_{i=2,4,6}\psi^*_if_i + \mathcal{M}_1^*,\\
	\mathcal{B}^* &= \bm{u}^* + \sum_{i=2,4,6}\bm{x}^*_ie_i
	+ \sum_{i=2,4,6}\bm{y}^*_if_i + \mathcal{B}_1^*,
\end{split}\]
where
\[
	\theta^* = \vartheta + \sum_{i=0}^6\theta^*_i\alpha^{\vee}_i,\quad
	\bm{u}^* = \sum_{i=0}^6\bm{u}^*_i\alpha^{\vee}_i.
\]

We next consider a gauge transformation
\[
	\mathcal{M}^{+} = \exp(\mathrm{ad}(\Gamma_2))\mathcal{M}^*,\quad
	d_{\bm{t}}-\mathcal{B}^{+}
	= \exp(\mathrm{ad}(\Gamma_2))(d_{\bm{t}}-\mathcal{B}^*),
\]
with $\Gamma_2=\sum_{i=2,4,6}\lambda_if_i$ such that
$\mathcal{M}^{+},\mathcal{B}^{+}\in\mathfrak{b}_{+}$, namely
\begin{equation}\label{Gauge_Trf_1}
	\xi^*_i\lambda_i^2 - (\theta^*|\alpha^{\vee}_i)\lambda_i
	- \psi^*_i = 0\quad (i=2,4,6),
\end{equation}
and
\begin{equation}\label{Gauge_Trf_2}
	d_{\bm{t}}\lambda_i = \bm{x}^*_i\lambda_i^2
	- (\bm{u}^*|\alpha^{\vee}_i)\lambda_i - \bm{y}^*_i\quad (i=2,4,6).
\end{equation}
Here we have
\begin{lem}\label{Lem:Borel}
Under the system \eqref{Sim_Red_ED_2}, the equation \eqref{Gauge_Trf_2}
follows from the equation \eqref{Gauge_Trf_1}.
\end{lem}
\begin{proof}
The system \eqref{Sim_Red_ED_2} can be expressed as
\begin{equation}\begin{split}\label{Sim_Red_ED_3}
	&d_{\bm{t}}\theta^*_i = \bm{x}^*_i\psi^*_i - \bm{y}^*_i\xi^*_i,\quad
	d_{\bm{t}}\theta^*_j = 0,\\
	&d_{\bm{t}}\xi^*_i = (\bm{u}^*|\alpha^{\vee}_i)\xi^*_i
	- \bm{x}^*_i(\theta^*|\alpha^{\vee}_i),\\
	&d_{\bm{t}}\psi^*_i = -(\bm{u}^*|\alpha^{\vee}_i)\psi^*_i
	+ \bm{y}^*_i(\theta^*|\alpha^{\vee}_i),
\end{split}\end{equation}
for $i=2,4,6$ and $j=0,1,3,5$.
By using \eqref{Sim_Red_ED_3} and
$(d_{\bm{t}}\theta^*|\alpha^{\vee}_i)=2d_{\bm{t}}\theta^*_i$, we obtain
\[\begin{split}
	&d_{\bm{t}}\left(\xi^*_i\lambda_i^2-(\theta^*|\alpha^{\vee}_i)
	\lambda_i-\psi^*_i\right)\\
	&= \left\{2\xi^*_i\lambda_i-(\theta^*|\alpha^{\vee}_i)\right\}
	\left\{d_{\bm{t}}\lambda_i-\bm{x}^*_i\lambda_i^2
	+(\bm{u}^*|\alpha^{\vee}_i)\lambda_i+\bm{y}^*_i\right\}\quad
	(i=2,4,6).
\end{split}\]
It follows that the equation \eqref{Gauge_Trf_1} implies \eqref{Gauge_Trf_2}
or
\begin{equation}\label{Prf:Lem:Borel_1}
	\lambda_i = \frac{(\theta^*|\alpha^{\vee}_i)}{2\xi^*_i}\quad (i=2,4,6).
\end{equation}
Hence it is enough to verify that the equation \eqref{Gauge_Trf_2} follows
from \eqref{Prf:Lem:Borel_1}.
Together with \eqref{Sim_Red_ED_3}, the equation \eqref{Prf:Lem:Borel_1}
implies
\begin{equation}\begin{split}\label{Prf:Lem:Borel_2}
	d_{\bm{t}}\lambda_i &= \frac{(d_{\bm{t}}\theta^*|\alpha^{\vee}_i)\xi^*_i
	-(\theta^*|\alpha^{\vee}_i)d_{\bm{t}}\xi^*_i}{2(\xi^*_i)^2}\\
	&= \bm{x}^*_i\lambda_i^2 - (\bm{u}^*|\alpha^{\vee}_i)\lambda_i
	- \bm{y}^*_i + \frac{\bm{x}^*_i\{4\xi^*_i\psi^*_i
	+(\theta^*|\alpha^{\vee}_i)^2\}}{4(\xi^*_i)^2}.
\end{split}\end{equation}
On the other hand, we obtain
\begin{equation}\label{Prf:Lem:Borel_3}
	4\xi^*_i\psi^*_i + (\theta^*|\alpha^{\vee}_i)^2 = 0,
\end{equation}
by substituting \eqref{Prf:Lem:Borel_1} into \eqref{Gauge_Trf_1}.
Combining \eqref{Prf:Lem:Borel_2} and \eqref{Prf:Lem:Borel_3}, we obtain the
equation \eqref{Gauge_Trf_2}.
\end{proof}

Thanks to Lemma \ref{Lem:Borel}, the gauge parameters $\lambda_i$ $(i=2,4,6)$
are determined by the equation \eqref{Gauge_Trf_1}.
Hence we obtain the system on $\mathfrak{b}_{+}$
\begin{equation}\label{Sim_Red_Borel}
	d_{\bm{t}}\mathcal{M}^{+} = [\mathcal{B}^{+},\mathcal{M}^{+}],\quad
	d_{\bm{t}}\mathcal{B}^{+} = \mathcal{B}^{+}\wedge\mathcal{B}^{+},
\end{equation}
with dependent variables $\lambda_i$ and $\mu_i=\xi^*_i$ $(i=2,4,6)$.
The operator $\mathcal{M}^{+}$ is described as
\[\begin{split}
	\mathcal{M}^{+} &= \kappa + \sum_{i=2,4,6}\mu_ie_i + (c_0+\lambda_6)e_0
	+ (c_1+\lambda_2)e_1\\
	&\quad + (c_3+\lambda_2+\lambda_4+c_{63}\lambda_6-\lambda_2\lambda_4)e_3
	+ (c_5+\lambda_4)e_5 + e_{21}\\
	&\quad + e_{45} + e_{60} + (1-\lambda_4)e_{23} + (1-\lambda_2)e_{43}
	+ c_{63}e_{63} + e_{234},
\end{split}\]
where $\kappa\in\mathfrak{h}$.
Note that $d_{\bm{t}}\kappa=0$.

Let $s_1$ and $s_2$ be independent variables defined by
\[
	s_1 = \frac{c_{63}(1+c_3-c_0c_{63})}{6},\quad
	s_2 = \frac{c_{63}(1+c_1)(1+c_5)}{6}.
\]
We now regard the system \eqref{Sim_Red_Borel} as a system of ordinary
differential equations
\begin{equation}\label{Sim_Red_ODE}
	\left[s(s-1)\frac{d}{ds}-B,\mathcal{M}^{+}\right] = 0,
\end{equation}
with respect to the independent variable $s=s_1$ by setting $s_2=1$.
The operator $B$ is expressed in the form
\[\begin{split}
	B &= \sum_{i=0}^{6}u_i\alpha^{\vee}_i + \sum_{i=0}^{6}x_ie_i
	+ x_{21}e_{21} + x_{45}e_{45}+ x_{23}e_{23} + x_{43}e_{43}\\
	&\quad + x_{63}e_{63} + x_{234}e_{234} + x_{236}e_{236} + x_{436}e_{436}
	+ x_{6234}e_{6234}.
\end{split}\]
Each coefficient of $B$ is determined as a polynomial in $\lambda_i$ and
$\mu_i$; we do not give the explicit formula.

Let $q_i$, $p_i$ $(i=1,2,3)$ be dependent variables defined by
\begin{equation}\begin{split}\label{Trf:dep_Var}
	&q_1 = \frac{1-\lambda_2}{1+c_1},\quad
	q_2 = \frac{1-\lambda_4}{1+c_5},\quad
	q_3 = \frac{1+c_3-c_0c_{63}}{1+c_3+c_{63}\lambda_6},\\
	&p_1 = -\frac{(1+c_1)\mu_2}{6},\quad
	p_2 = -\frac{(1+c_5)\mu_4}{6},\\
	&p_3 = -\frac{(1+c_3+c_{63}\lambda_6)
	\{(1+c_3+c_{63}\lambda_6)\mu_6+c_{63}(\kappa|\alpha_6^{\vee})\}}
	{6c_{63}(1+c_3-c_0c_{63})}.
\end{split}\end{equation}
We also set
\[
	\alpha_i = \frac{(\kappa|\alpha^{\vee}_i)}{6}\quad (i=0,\ldots,6).
\]
Then we obtain
\begin{thm}
The system \eqref{Sim_Red_ODE} is equivalent to the system \eqref{Sys:E6} with
\eqref{Sys:E6_Ham}.
\end{thm}

In the last, we note a derivation of the affine Weyl group symmetry for the
system \eqref{Sys:E6}.
We define a Poisson structure for the $\mathfrak{b}_{+}$-valued operator
$\mathcal{M}^{+}$ by
\[
	\{\mu_i,\lambda_j\} = 6\delta_{i,j},\quad
	\{\mu_i,\mu_j\} = \{\lambda_i,\lambda_j\} = 0\quad (i,j=2,4,6).
\]
It is equivalent to
\[
	\{p_i,q_j\} = \delta_{i,j},\quad \{p_i,p_j\} = \{q_i,q_j\} = 0\quad
	(i,j=1,2,3),
\]
via the transformation \eqref{Trf:dep_Var}.
Hence $p_i$, $q_i$ $(i=1,2,3)$ give a canonical coordinate system associated
with the Poisson structure for $\mathcal{M}^{+}$.

Thanks to \cite{NY2}, we then obtain a birational canonical transformations
$r_i$ $(i=0,\ldots,6)$ given in Theorem \ref{Bac_Trf}.
They are derived from the transformations
\[
	r_i(X) = X\exp(-e_i)\exp(f_i)\exp(-e_i)\quad (i=0,\ldots,6),
\]
where $X=\exp(\sum_{k\in\mathbb{N}}t_k\Lambda_k)X(0)$.

\section*{Acknowledgement}
The authers are grateful to Professors Masatoshi Noumi, Yasuhiko Yamada and Teruhisa Tsuda for valuable discussions and advices.


\end{document}